\documentclass[12pt,showpacs,aps,prd]{revtex4}
\usepackage{graphicx}
\input epsf

\begin{document}

\title{Could $Z_{c}(3900)$ be a $I^{G}J^{P}=1^{+}1^{+}$ $D^{*}\bar{D}$ molecular state?}
\author{Chun-Yu Cui$^{\P}$, Xin-Hua Liao$^{\P}$, Yong-Lu Liu$^*$, and Ming-Qiu Huang$^*$}
\affiliation{$^{\P}$ Department of Physics, School of Biomedical Engineering, Third Military Medical University, Chongqing 400038, China\\$^*$ Department of Physics, National University of Defense
Technology, Hunan 410073, China}
\date{\today}
\begin{abstract}
We investigate the nature of the recently observed narrow resonance $Z_{c}(3900)$, which is assumed to be a
$D^{*}\bar{D}$ molecular state with quantum numbers $I^{G}J^{P}=1^{+}1^{+}$. Using QCD sum rules, we consider contributions up to dimension eight in the operator product expansion and work at the leading order in $\alpha_{s}$. The mass we arrived at is $(3.88 \pm 0.17)~\mbox{GeV}$, which coincides with the mass of $Z_{c}(3900)$.
\end{abstract}
\maketitle

\section{Introduction}\label{sec1}
The charmonium-like exotic states are of great interest as they provide a satisfied window both in studying the dynamics of light quarks interact with heavy quarks and testing the standard model itself. In the past few years, many exotic states have been observed by the collaborations such as BEC, BELLE, BABAR, CDF and D0. They are also investigated with various models theoretically. These exotic states can be illustrated either as molecular states or multi-quark states or hybrids(see reviews \cite{Swanson:2006st}-\cite{Nora} and references therein). Although there is even no one conformed conclusion on these configurations, it is still interesting to investigate the inner structure of these states.

Recently, the BECIII Collaboration~\cite{BEC} reported a new enhancement structure $Z_{c}(3900)$ in the $\pi^{\pm}J/\Psi$ invariant mass spectrum of the $Y(4260)\rightarrow J/\Psi\pi^+\pi^-$ decay.
The mass and width of this state is $M=(3899.0\pm 3.6\pm 4.9)~{\rm MeV}/c^2$ and $\Gamma=(46\pm 10\pm 20)$~MeV/$c^{2}$. BELLE confirmed this observation with mass $M=(3894.5\pm 6.6\pm 4.5)~{\rm
MeV}/c^2$ and width $\Gamma=(63\pm 24\pm 26)$~MeV/$c^{2}$, where the
errors are statistical and systematic, respectively~\cite{BELLE}. Before this observation, its existence is predicted by the ISPE mechanism~\cite{Liu} and in the molecular and tetraquark schemes, respectively~\cite{Zhang,Maiani}. After the new experimental observation, there are many investigations about its possible internal configuration, such as molecular states~\cite{Zhao,Ke:2013gia,Dong:2013iqa,Wilbring:2013cha,Guo}, tetraquark states~\cite{Maiani2,Voloshin,Dias:2013xfa,Braaten:2013boa,Qiao:2013raa,Wang}, the re-scattering effects~\cite{Liu:2013vfa,Li}, and so on~\cite{Chen:2013coa,Liu:2013}. In Ref.~\cite{Zhao}, they give an explanation of Y(4260) as a ${\bar D} D_{1}(2420)+{\bar D} D_{1}(2420)$ and interpret $Z_{c}(3900)$ as a $D^{*}{\bar D}$ bound state. Based on heavy quark spin symmetry and heavy flavour symmetry, the authors in~\cite{Guo} predict $Z_{c}(3900)$ as the isovector $D^{*}{\bar D}$ partners of the $Z_{b}(10610)$.
In Ref.~\cite{Maiani2}, the author discuss the $J^{PG}=1^{++}$ $Z_{c}(3900)$ within
tetraquark model as well as molecular model, then investigate its various decay modes. In Ref.~\cite{Voloshin}, $Z_{c}(3900)$ is studied in various models(the molecular as well as the hadro-charmonium and tetraquark schemes), and by the ways of distinguishing for further experimental studies. However, it is argued that the molecular
interpretation is less likely and this could be investigated with QCD sum rules~\cite{Namit}.

Due to the asymptotic property of the QCD, study of the hadron spectrum have to concern about the nonperturbative effect which is difficult in quantum field theory. There are many methods to estimate the mass of a hadron, among which QCD sum rule(QCDSR)~\cite{svz,reinders,overview2,overview3,NielsenPR} is a fairly reliable one. Quantum numbers compatible with the experiment are the fundamental ingredients in QCDSR analysis of composite particles. Since $Z_{c}(3900)$ was observed in the $Y(4260)\rightarrow Z_{c}(3900)\pi$ decay process, the assignment of $I^{G}=1^{+}$ is known. In this article, by assuming $Z_{c}(3900)$ as a $D^{-}D^{*0}$ molecular state with $I^{G}J^{P}=1^{+}1^{+}$, we investigate the mass of this possible molecular configuration within the framework of QCD sum rules. We construct the following interpolating current to represent the $Z_{c}(3900)$ molecular state
\begin{eqnarray}
j^{\mu}&=&\frac{1}{\sqrt{2}}[(\bar{u}i\gamma^{5}c)(\bar{c}\gamma_{\mu}d)+(\bar{u}\gamma_{\mu}c)(\bar{c}i\gamma^{5}d)].
\end{eqnarray}

The rest of the paper is organized as three parts. The QCDSR for the $Z_{c}(3900)$ is derived in Sec. \ref{sec2}, with contributions up to dimension eight in the operator product expansion(OPE). The numerical analysis is presented to extract the hadronic mass at the end of this section. Sec. \ref{sec3} is the summary and conclusion.

\section{QCD sum rules for $Z_{c}(3900)$}\label{sec2}
In the QCDSR approach, the mass of the particle can be determined by considering the two-point correlation function
\begin{eqnarray}
\Pi^{\mu\nu}(q^{2})=i\int
d^{4}x\mbox{e}^{iq.x}\langle0|T[j^{\mu}(x)j^{\nu+}(0)]|0\rangle.
\end{eqnarray}
Lorentz covariance implies that the two-point correlation function can be generally parameterized as
\begin{eqnarray}
\Pi^{\mu\nu}(q^{2})=(\frac{q^{\mu}q^{\nu}}{q^{2}}-g^{\mu\nu})\Pi^{(1)}(q^{2})+\frac{q^{\mu}q^{\nu}}{q^{2}}\Pi^{(0)}(q^{2}).
\end{eqnarray}
We select the term proportional to $g_{\mu\nu}$ to extract the mass sum rule, since it gets contributions only from the $1^{+}$ state. The QCD sum rule attempts to link the hadron phenomenology with the interactions of quarks and gluons. It contains three main ingredients: an approximate description of the correlation function in terms
of intermediate states through the dispersion relation, a evaluation of the same correlation function in terms of QCD degrees of freedom via an OPE, and a procedure for matching these two descriptions and extracting the parameters that characterize the hadronic state of interest.

We can insert a complete set of intermediate hadronic states with the same quantum numbers
as the current operators $j^{\mu}$ into the correlation function to obtain the
phenomenological side. The coupling of the current with the state can be defined by the coupling constant as follows:
\begin{eqnarray}
\langle 0|j_{\mu}|Z\rangle&=&\lambda \epsilon_{\mu}.
\end{eqnarray}
Phenomenologically, $\Pi^{(1)}(q^{2})$ can be expressed as
\begin{eqnarray}\label{ph}
\Pi^{(1)}(q^{2})=\frac{\lambda^{2}}{M_{Z}^{2}-q^{2}}+\frac{1}{\pi}\int_{s_{0}}
^{\infty}ds\frac{\mbox{Im}\Pi^{(1)\mbox{phen}}(s)}{s-q^{2}},
\end{eqnarray}
where $M_{Z}$ denotes the mass of the molecular state, and $s_0$ is the continuum threshold parameter.

In the OPE side, $\Pi^{(1)}(q^{2})$ can be written as
\begin{eqnarray}\label{ope}
\Pi^{(1)}(q^{2})=\int_{4m_{c}^{2}}^{\infty}ds\frac{\rho^{\mbox{OPE}}(s)}{s-q^{2}}+\Pi_{1}^{\mbox{cond}}(q^{2}),
\end{eqnarray}
where the spectral density is $\rho^{OPE}(s)=\frac{1}{\pi}\mbox{Im}\Pi^{\mbox{(1)}}(s)$.
Applying the quark-hadron duality hypothesis with the Borel transformation, one obtains the following sum rule:
\begin{eqnarray}\label{sr}
\lambda^{2}e^{-M_{Z}^{2}/M^{2}}&=&\int_{4m_{c}^{2}}^{s_{0}}ds\rho^{\mbox{OPE}}e^{-s/M^{2}}+\hat{B}\Pi_{1}^{\mbox{cond}},
\end{eqnarray}
with $M^2$ the Borel parameter.

In the OPE side, we work at the leading order in $\alpha_{s}$
and consider vacuum condensates up to dimension eight, with the similar
techniques in Refs.~\cite{technique}. In order to consider the isospin violation, we keep the terms which are linear in the light-quark masses $m_{u}$ and $m_{d}$. After some tedious OPE calculations, the concrete forms of spectral densities can be derived:
\begin{eqnarray}
\rho^{OPE}(s)=\rho^{\mbox{pert}}(s)+\rho^{\langle\bar{q}q\rangle}(s)+\rho^{\langle
g^{2}G^{2}\rangle}(s)+\rho^{\langle
g\bar{q}\sigma\cdot G q\rangle}(s)+\rho^{\langle\bar{q}q\rangle^{2}}(s)+\rho^{\langle g^{3}G^{3}\rangle}(s)+\rho^{\langle\bar{q}q\rangle\langle
g^{2}G^{2}\rangle},
\end{eqnarray}
with
\begin{eqnarray}\label{spectralmole}
\rho^{\mbox{pert}}(s)&=&\frac{3}{2^{12}\pi^{6}}\int_{\alpha_{min}}^{\alpha_{max}}\frac{d\alpha}{\alpha^{3}}\int_{\beta_{min}}^{1-\alpha}\frac{d\beta}{\beta^{3}}(1-\alpha-\beta)(1+\alpha+\beta)r(m_{c},s)^{4}
\nonumber\\&&{}
+\frac{3m_{c}}{2^{11}\pi^{6}}\int_{\alpha_{min}}^{\alpha_{max}}\frac{d\alpha}{\alpha^{3}}\int_{\beta_{min}}^{1-\alpha}\frac{d\beta}{\beta^{3}}(\alpha+\beta-1)(m_{u}\alpha^2+m_{d}\beta^2
\nonumber\\&&{}
+m_{u}\alpha\beta+m_{d}\alpha\beta+3m_{u}\alpha+3m_{d}\beta)r(m_{c},s)^{3}
,\nonumber\\
\rho^{\langle\bar{q}q\rangle}(s)&=&-\frac{3\langle\bar{q}q\rangle}{2^{8}\pi^{4}}m_{c}\int_{\alpha_{min}}^{\alpha_{max}}\frac{d\alpha}{\alpha^{2}}\int_{\beta_{min}}^{1-\alpha}\frac{d\beta}{\beta^{2}}(\alpha+\beta)(1+\alpha+\beta)r(m_{c},s)^{2}
\nonumber\\&&{}
+\frac{3\langle\bar{q}q\rangle}{2^{8}\pi^{4}}(m_{u}+m_{d})\int_{\alpha_{min}}^{\alpha_{max}}\frac{d\alpha}{\alpha(1-\alpha)}[m_{c}^{2}-\alpha(1-\alpha)s]^2
\nonumber\\&&{}
+\frac{3\langle\bar{q}q\rangle}{2^{6}\pi^{4}}m_{c}^2(m_{u}+m_{d})\int_{\alpha_{min}}^{\alpha_{max}}\frac{d\alpha}{\alpha}\int_{\beta_{min}}^{1-\alpha}\frac{d\beta}{\beta}r(m_{c},s)
\nonumber\\&&{}
-\frac{3\langle\bar{q}q\rangle}{2^{8}\pi^{4}}(m_{u}+m_{d})\int_{\alpha_{min}}^{\alpha_{max}}\frac{d\alpha}{\alpha}\int_{\beta_{min}}^{1-\alpha}\frac{d\beta}{\beta}r(m_{c},s)^2
,\nonumber\\
\rho^{\langle g^{2}G^{2}\rangle}(s)&=&\frac{\langle
g^{2}G^{2}\rangle}{2^{11}\pi^{6}}m_{c}^{2}\int_{\alpha_{min}}^{\alpha_{max}}d\alpha\int_{\beta_{min}}^{1-\alpha}\frac{d\beta}{\beta^{3}}(1-\alpha-\beta)(1+\alpha+\beta)r(m_{c},s)\nonumber\\&&{}
+\frac{\langle
g^{2}G^{2}\rangle}{2^{11}\pi^{6}}\int_{\alpha_{min}}^{\alpha_{max}}\frac{d\alpha}{\alpha}\int_{\beta_{min}}^{1-\alpha}\frac{d\beta}{\beta^{2}}(2\alpha+2\beta-1)r(m_{c},s)^{2}
,\nonumber\\
\rho^{\langle g\bar{q}\sigma\cdot G q\rangle}(s)&=&-\frac{3\langle
g\bar{q}\sigma\cdot G
q\rangle}{2^{7}\pi^{4}}m_{c}\int_{\alpha_{min}}^{\alpha_{max}}\frac{d\alpha}{\alpha}[m_{c}^{2}-\alpha(1-\alpha)s]
\nonumber\\&&{}
+\frac{3\langle g\bar{q}\sigma\cdot
Gq\rangle}{2^{8}\pi^{4}}m_{c}\int_{\alpha_{min}}^{\alpha_{max}}d\alpha\int_{\beta_{min}}^{1-\alpha}\frac{d\beta}{\beta}r(m_{c},s)
,\nonumber\\&&{}
+\frac{3\langle g\bar{q}\sigma\cdot
Gq\rangle}{2^{7}\pi^{4}}m_{c}\int_{\alpha_{min}}^{\alpha_{max}}d\alpha\int_{\beta_{min}}^{1-\alpha}\frac{d\beta}{\beta^2}(\alpha+\beta)r(m_{c},s)
,\nonumber\\
\rho^{\langle\bar{q}q\rangle^{2}}(s)&=&\frac{\langle\bar{q}q\rangle^{2}}{2^{4}\pi^{2}}m_{c}^{2}\sqrt{1-4m_{c}^{2}/s}
,\nonumber\\
\rho^{\langle g^{3}G^{3}\rangle}(s)&=&\frac{\langle
g^{3}G^{3}\rangle}{2^{12}\pi^{6}}m_{c}^{2}\int_{\alpha_{min}}^{\alpha_{max}}d\alpha\int_{\beta_{min}}^{1-\alpha}\frac{d\beta}{\beta^{3}}\alpha(1-\alpha-\beta)(1+\alpha+\beta)
\nonumber\\&&{}
+\frac{\langle
g^{3}G^{3}\rangle}{2^{13}\pi^{6}}\int_{\alpha_{min}}^{\alpha_{max}}d\alpha\int_{\beta_{min}}^{1-\alpha}\frac{d\beta}{\beta^{3}}(1-\alpha-\beta)(1+\alpha+\beta)r(m_{c},s)
,\nonumber\\
\rho^{\langle\bar{q}q\rangle\langle
g^{2}G^{2}\rangle}(s)&=&\frac{\langle\bar{q}q\rangle\langle
g^{2}G^{2}\rangle}{3\cdot2^{9}\pi^{4}}m_{c}\bigg[-2\sqrt{1-\frac{4m_{c}^{2}}{s}}+\int_{\alpha_{min}}^{\alpha_{max}}d\alpha\int_{\beta_{min}}^{1-\alpha}\frac{d\beta}{\beta^{2}}(\beta^2-3\alpha^2-3\alpha\beta-3\alpha)\bigg],\nonumber
\end{eqnarray}
with $r(m_{c},s)=(\alpha+\beta)m_{c}^2-\alpha\beta s$. The integration limits are given by $\alpha_{min}=\Big(1-\sqrt{1-4m_{c}^{2}/s}\Big)/2$, $\alpha_{max}=\Big(1+\sqrt{1-4m_{c}^{2}/s}\Big)/2$, and $\beta_{min}=\alpha m_{c}^{2}/(s\alpha-m_{c}^{2})$.
The term $\hat{B}\Pi_{1}^{\mbox{cond}}$ reads
\begin{eqnarray}
\hat{B}\Pi_{1}^{\mbox{cond}}&=&\frac{\langle\bar{q}q\rangle\langle
g^{2}G^{2}\rangle}{3\cdot2^{9}\pi^{4}}m_{c}^{3}\int_{0}^{1}d\alpha\int_{0}^{1-\alpha}\frac{d\beta}{\beta^3}(\alpha+\beta)(1+\alpha+\beta)e^{-\frac{(\alpha+\beta)m_{c}^{2}}{\alpha\beta M^{2}}}
\nonumber\\&&
-\frac{\langle g^{2}G^{2}\rangle^{2}}{3^{2}\cdot2^{15}\pi^{6}}m_{c}^{4}\int_{0}^{1}\frac{d\alpha}{\alpha^{2}}\int_{0}^{1-\alpha}\frac{d\beta}{\beta^{2}}(\alpha+\beta-1)(1+\alpha+\beta)\frac{1}{M^{2}}e^{-\frac{(\alpha+\beta)m_{c}^{2}}{\alpha\beta M^{2}}}\nonumber\\&&
-\frac{\langle g^{2}G^{2}\rangle^{2}}{3^{2}\cdot2^{13}\pi^{6}}m_{c}^{2}\int_{0}^{1}\frac{d\alpha}{\alpha^{2}}\int_{0}^{1-\alpha}d\beta(2\alpha+2\beta-1)e^{-\frac{(\alpha+\beta)m_{c}^{2}}{\alpha\beta M^{2}}}.\nonumber
\end{eqnarray}

Taking the derivative of Eq.(\ref{sr}) with respect to $\frac{1}{M^2}$ and then dividing by itself, we extract the molecular state mass
\begin{eqnarray}\label{sumrule}
M_{Z}^{2}&=&\bigg\{\int_{4m_{c}^{2}}^{s_{0}}ds\rho^{\mbox{OPE}}s
e^{-s/M^{2}}+\frac{d\hat{B}\Pi_{1}^{\mbox{cond}}}{d(-\frac{1}{M^{2}})}\bigg\}/
\bigg\{\int_{4m_{c}^{2}}^{s_{0}}ds\rho^{\mbox{OPE}}e^{-s/M^{2}}+\hat{B}\Pi_{1}^{\mbox{cond}}\bigg\}
\end{eqnarray}

For numerical analysis of the equation (\ref{sumrule}), we first specify the input parameters. The quark masses are taken as $m_{u}=2.3~\mbox{MeV}$ and $m_{d}=4.8~\mbox{MeV}$\cite{PDG}. The condensates are $\langle\bar{u}u\rangle=\langle\bar{d}d\rangle=\langle\bar{q}q\rangle=-(0.23\pm0.03)^{3}~\mbox{GeV}^{3}$,
$\langle g\bar{q}\sigma\cdot G
q\rangle=m_{0}^{2}~\langle\bar{q}q\rangle$,
$m_{0}^{2}=0.8~\mbox{GeV}^{2}$, $\langle
g^{2}G^{2}\rangle=0.88~\mbox{GeV}^{4}$, and $\langle
g^{3}G^{3}\rangle=0.045~\mbox{GeV}^{6}$~\cite{overview2}.
Complying with the standard procedure of the sum rule analysis, the threshold $s_{0}$ and Borel parameter $M^{2}$ are varied to find the optimal stability window. There are two criteria (pole dominance and convergence of the OPE) for choosing the Borel parameter $M^{2}$ and threshold $s_{0}$.

\begin{figure}
\centering
\includegraphics[totalheight=5cm,width=7cm]{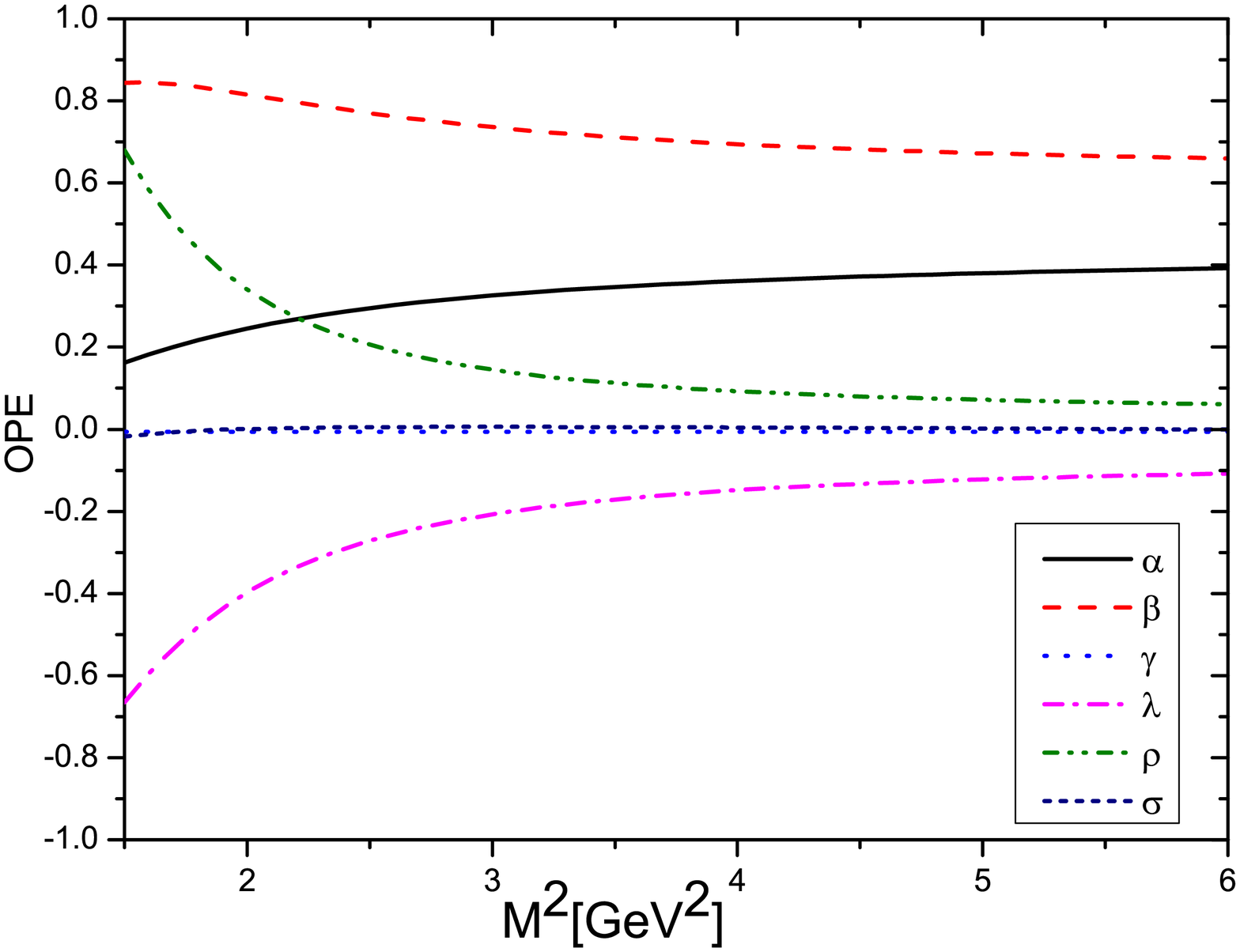}
\includegraphics[totalheight=5cm,width=7cm]{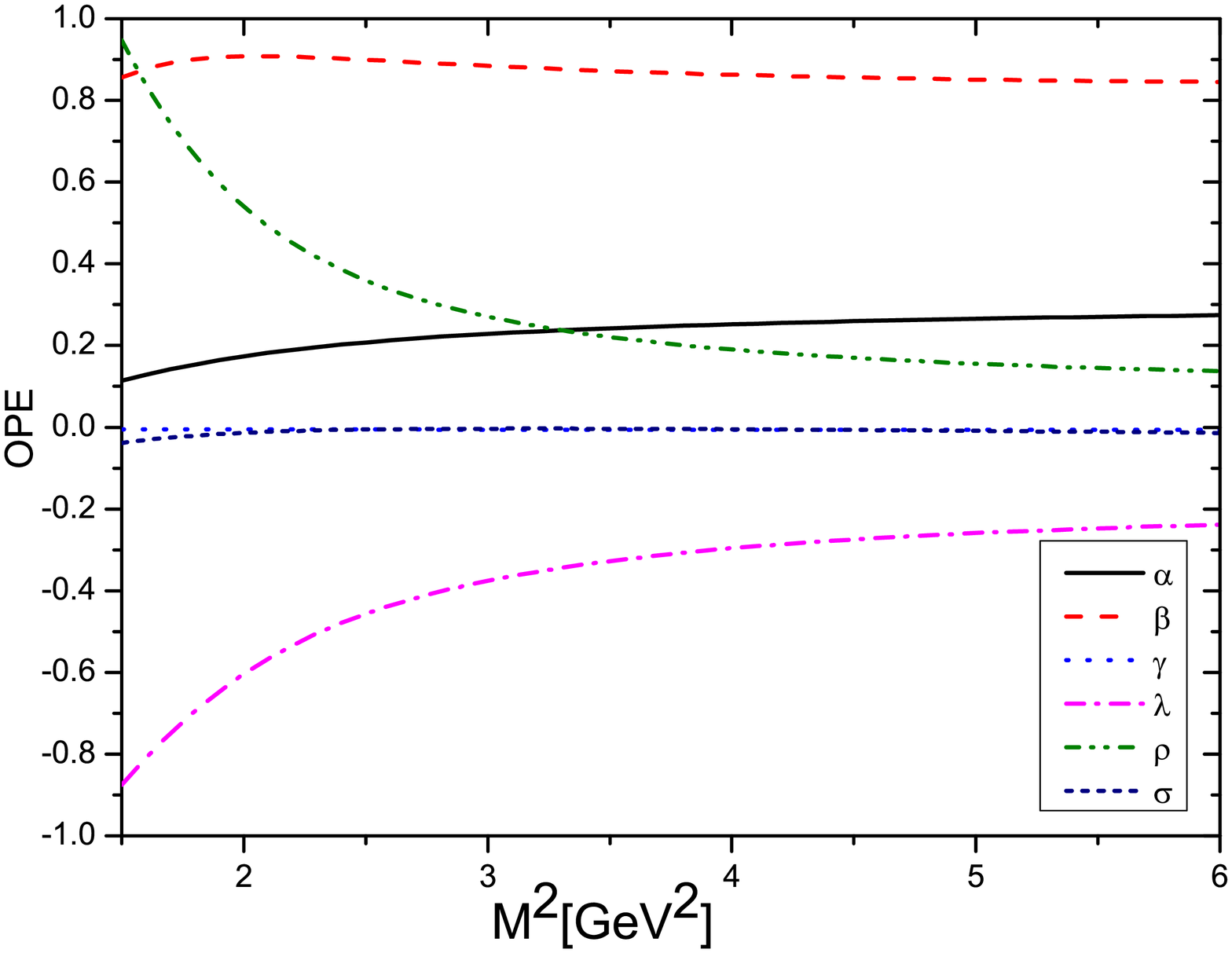}
\caption{a) The OPE convergence for the molecular state for $m_{c}=1.275\,\rm{GeV}$. The contributions from different terms with variation of the Borel parameter $M^2$ in the OPE. The notations $\alpha$, $\beta$, $\gamma$, $\lambda$ $\rho$ and $\sigma$ correspond to perturbative, $D=3$ two-quark, $D=4$ two-gluon, $D=5$ mixed, $D=6$ four-quark plus three-gluon, $D=7$ two-quark multiply two-gluon contributions, respectively; b) The same as a) but for $m_{c}=1.47\,\rm{GeV}$}\label{fig1}
\end{figure}

We take QCDSR analysis in both case of $\overline{MS}$ mass $m_{c}=1.275\,\rm{GeV}$ and pole mass $m_{c}=1.47\,\rm{GeV}$~\cite{PDG}.
The contributions from various terms in the OPE are shown in Fig.\ref{fig1}. The dimension-$8$ condensate term is rather small compared with the total contributions, thus, we omit it for clarity. We have used $\sqrt{s_0}= 4.4\,\mbox{GeV}$. We notice that two-quark condensate is very large and plays a dominant role in the OPE side. Luckily, $\langle\bar{q}q\rangle$ and $\langle g\bar{q}\sigma\cdot G q\rangle$ condensates have different signs and they could cancel out each other to a big extent. When $m_{c}=1.275\,\rm{GeV}$, from Fig.\ref{fig1}a) it can be seen that for $M^2\geq 2.50\,\mbox{GeV}^2$, the contribution of the dimension-$6$ condensate is less than $20\%$ of the total contribution and  the contribution of the dimension-$5$ condensate is less than $25\%$ of the total contribution, which indicate the starting point for a good Borel convergence. Therefore, we fix the uniform lower value of $M^2$ in the sum rule window as $M^2_{min}= 2.50\,\mbox{GeV}^2$. Fig.\ref{fig1}b) shows a slow convergence of the OPE. When $M^2\geq 3.0\,\mbox{GeV}^2$, the contribution of the dimension-$6$ condensate is less than $28\%$ of the total contribution and  the contribution of the dimension-$5$ condensate is less than $35\%$ of the total contribution.

\begin{figure}
\centering
\includegraphics[totalheight=5cm,width=7cm]{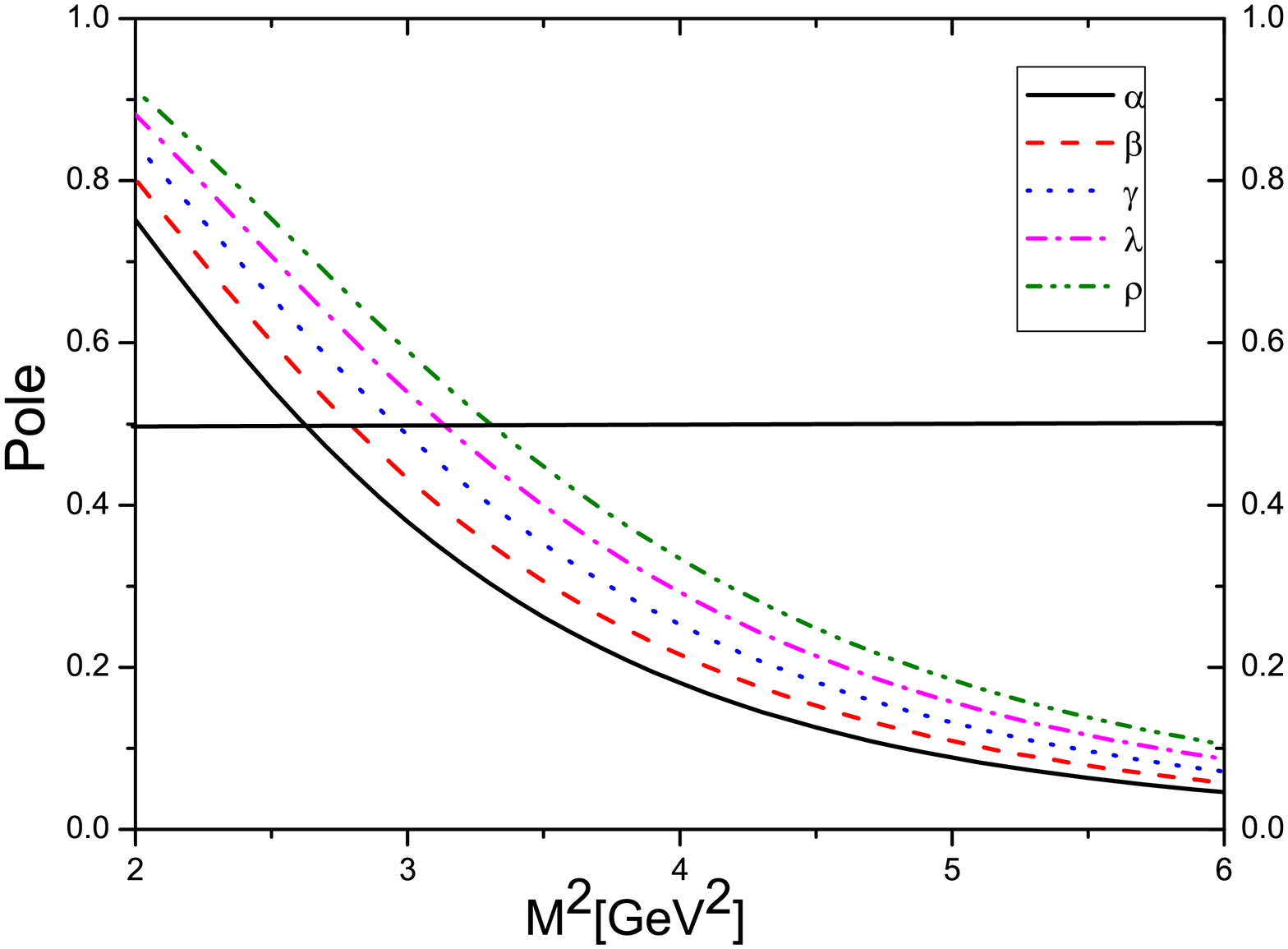}
\includegraphics[totalheight=5cm,width=7cm]{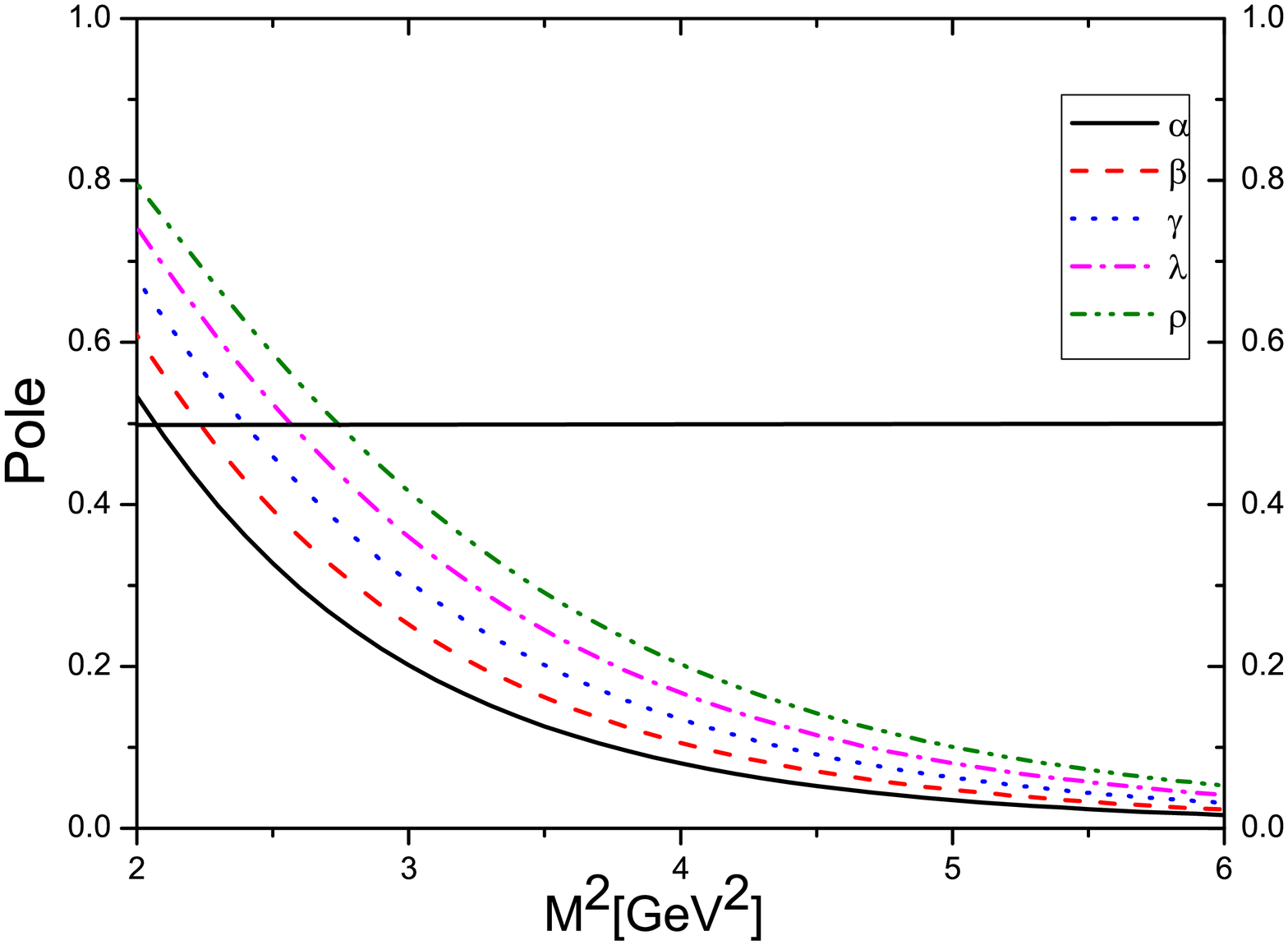}
\caption{a) The contributions from the pole terms with variation of the Borel parameter $M^2$ in the case of molecular state for $m_{c}=1.275\,\rm{GeV}$. The notations $\alpha$, $\beta$, $\gamma$, $\lambda$ and $\rho$  correspond to the threshold parameters $\sqrt{s_0}=4.2\,\rm{GeV}$, $4.3\,\rm{GeV}$, $4.4\,\rm{GeV}$, $4.5\,\rm{GeV}$ and $4.6\,\rm{GeV}$, respectively; b) The same as a) but for $m_{c}=1.47\,\rm{GeV}$}\label{fig2}
\end{figure}

The upper limit of $M^2$ is determined by imposing that the pole contribution should be larger than the continuum contribution. Fig.\ref{fig2} shows that the contributions from the pole terms with variation of the Borel parameter $M^2$. We ask the pole contribution to be larger than $50\%$. We show in Table~\ref{tab:1} the values of $M^2_{max}$ for several values of $\sqrt{s_0}$. We observe that it is impossible to find a rational Borel window where both the pole dominance
and the OPE convergence satisfy well for $m_{c}=1.47\,\rm{GeV}$. 

\begin{figure}
\centerline{\epsfysize=6.0truecm\epsfbox{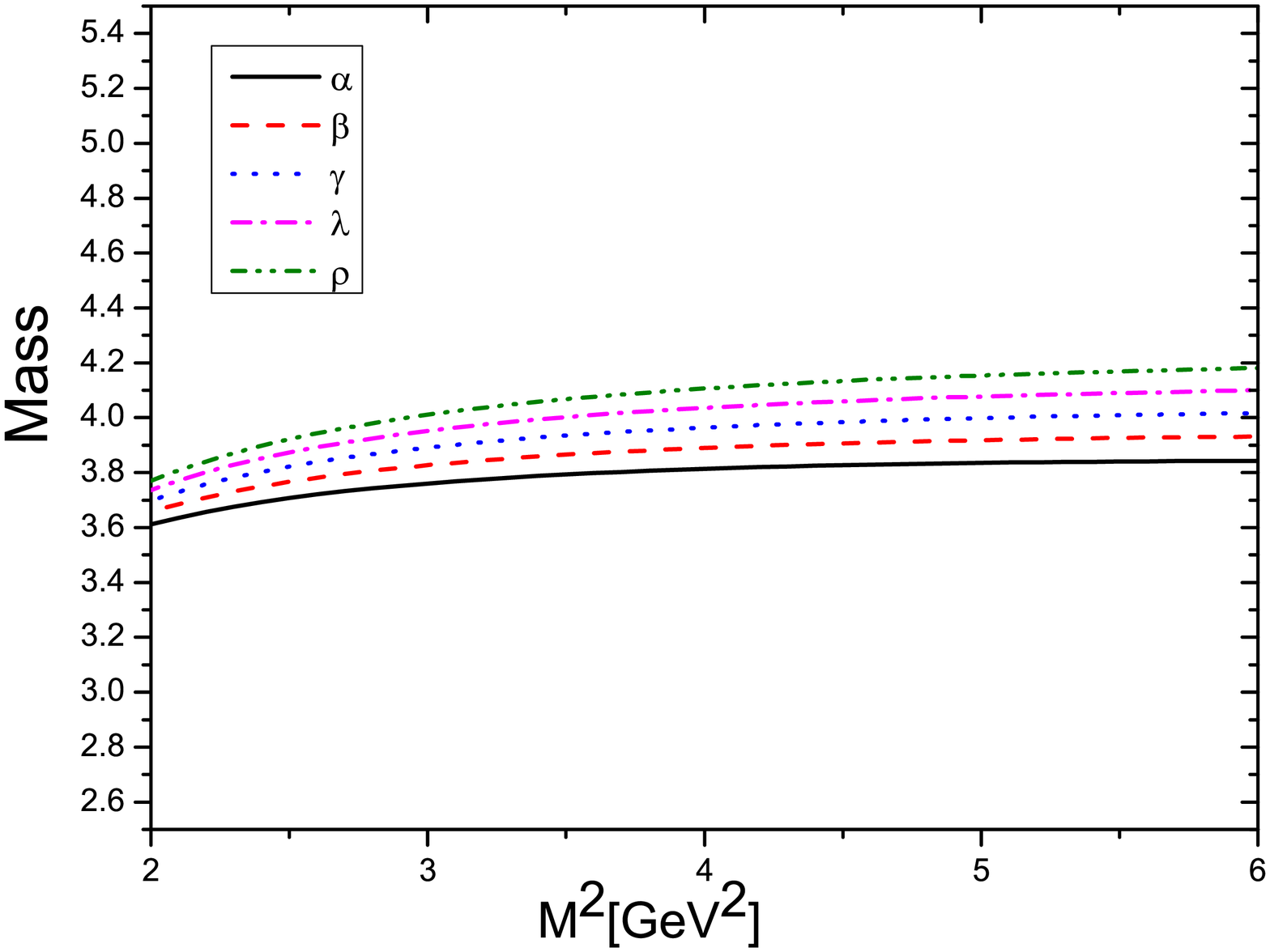}}
\caption{The mass of the molecular state as
a function of $M^2$ from sum rule (\ref{sumrule}) for $m_{c}=1.275\,\rm{GeV}$. The notations $\alpha$, $\beta$, $\gamma$, $\lambda$ and $\rho$ correspond to the threshold parameters $\sqrt{s_0}=4.2\,\rm{GeV}$, $4.3\,\rm{GeV}$, $4.4\,\rm{GeV}$, $4.5\,\rm{GeV}$ and $4.6\,\rm{GeV}$, respectively.}\label{fig3}
\end{figure}

Then, we focus on mass analysis for $m_{c}=1.275\,\rm{GeV}$. In Fig.\ref{fig3}, we show the molecular state mass, for different values of $\sqrt{s_0}$, in the relevant sum rule window. It can be seen that the mass is stable in the Borel window with the corresponding threshold $\sqrt{s_0}$. The final estimate of the $I^{G}J^{P}=1^{+}1^{+}$ molecular state is obtained as
\begin{eqnarray}
M_{Z} = (3.88 \pm 0.17)~\mbox{GeV}.
\label{Zmass}
\end{eqnarray}

\begin{table}
\caption{Upper limits in the Borel window for the $I^{G}J^{P}=1^{+}1^{+}$ $D^*\bar{D}$ current obtained from the sum rule for different values of $\sqrt{s_0}$.}\label{tab:1}
\begin{center}
\begin{tabular}{|c|c|c|}
\hline
$\sqrt{s_0}~(\mbox{GeV})$ & $M^2_{max}(\mbox{GeV}^2)$($m_{c}=1.275\,\rm{GeV}$)& $M^2_{max}(\mbox{GeV}^2)$($m_{c}=1.47\,\rm{GeV}$)\\ \hline
4.2 &2.7 &2.1\\ \hline
4.3 &2.8 &2.2\\ \hline
4.4 &3.0 &2.4\\ \hline
4.5 &3.2 &2.5\\ \hline
4.6 &3.4 &2.7\\ \hline
\end{tabular}
\end{center}
\end{table}

\section{Summary and conclusion}\label{sec3}
In summary, by assuming $Z_{c}(3900)$ as a $D^{*}\bar{D}$ molecular state with quantum numbers $I^{G}J^{P}=1^{+}1^{+}$, we have constructed and analyzed the
QCDSR to calculate the mass of the resonance.
Our numerical result is $M_{Z}=(3.88\pm0.17)~\mbox{GeV}$ for molecular state.
The result is compatible with the experimental data of $Z_{c}(3900)$ by the BEC and BELLE Collaborations.
Our finding indicates that $Z_{c}(3900)$ could be the isovector charmonium partners of the $Z_{b}(10610)$~\cite{10610}.
\begin{acknowledgments}
This work was supported in part by the National Natural Science
Foundation of China under Contract Nos.11347174, 11275268, 11105222 and 11025242.
\end{acknowledgments}

\end{document}